\documentclass[aps, prl, reprint, superscriptaddress]{revtex4-1}
\usepackage{graphicx}
\usepackage{amssymb,amsmath}
\usepackage{soul}
\usepackage{ulem}
\usepackage[font=scriptsize]{caption}
\usepackage{siunitx}
\usepackage[usenames,dvipsnames]{color}

\begin{document}
\title{Blind Ghost Imaging}

\author{A. M. Paniagua-Diaz}
\thanks{These authors contributed equally.}
\affiliation{University of Exeter, Stocker Road, Exeter EX4 4QL, United Kingdom}
\author{I. Starshynov}
\thanks{These authors contributed equally.}
\affiliation{University of Exeter, Stocker Road, Exeter EX4 4QL, United Kingdom}
\author{N. Fayard}
\thanks{These authors contributed equally.}
\affiliation{ESPCI Paris, PSL Research University, CNRS, Institut Langevin, 1 rue Jussieu, F-75005, Paris, France}
\author{A.~Goetschy}
\affiliation{ESPCI Paris, PSL Research University, CNRS, Institut Langevin, 1 rue Jussieu, F-75005, Paris, France}
\author{R. Pierrat}
\affiliation{ESPCI Paris, PSL Research University, CNRS, Institut Langevin, 1 rue Jussieu, F-75005, Paris, France}
\author{R. Carminati}
\email[Corresponding author:]{remi.carminati@espci.fr}
\affiliation{ESPCI Paris, PSL Research University, CNRS, Institut Langevin, 1 rue Jussieu, F-75005, Paris, France}
\author{J. Bertolotti}
\email[Corresponding author:]{j.bertolotti@exeter.ac.uk}
\affiliation{University of Exeter, Stocker Road, Exeter EX4 4QL, United Kingdom}

\begin{abstract}
Ghost imaging is an unconventional optical imaging technique that reconstructs the shape of an object combining the measurement of two signals: one that interacted with the object, but without any spatial information, the other containing spatial information, but that never interacted with the object~\cite{quantumghostimaging, classicghostimaging}. Ghost imaging is a very flexible technique, that has been generalized to the single-photon regime~\cite{singlephotghostimaging}, to the time domain~\cite{timeghostimaging}, to infrared and terahertz frequencies~\cite{thzghostimaging}, and many more conditions~\cite{introghostimaging}.
Here we demonstrate that ghost imaging can be performed without ever knowing the patterns illuminating the object, but using patterns correlated with them, doesn't matter how weakly. As an experimental proof we exploit the recently discovered correlation between the reflected and transmitted light from a scattering layer~\cite{CRtheory1,CRTcorr}, and reconstruct the image of an object hidden behind a scattering layer using only the reflected light, which never interacts with the object. This method opens new perspectives for non-invasive imaging behind or within turbid media.
\end{abstract}

\maketitle

In its simplest form ghost imaging (GI), also known as single pixel camera~\cite{singlepixelcamera} or dual photography~\cite{dualphotography}, is an imaging technique where, instead of illuminating uniformly an object and then detect the scattered light with a multipixel camera, the object is illuminated with a sequence of known patterns, and the scattered light is detected by a single photodiode~\cite{classicghostimaging}. 
By using enough illumination patterns, high quality images can be formed~\cite{hiresghostimaging}. Ghost imaging finds application in all cases where large arrays of detectors are harder to come by than reliable sources (e.g. THz imaging~\cite{thzghostimaging}), or when only a very small amount of signal is available~\cite{singlephotghostimaging}.
As there is a lot of freedom in the choice of the patterns used, one can optimize them to increase resolution in the areas of interest~\cite{foveatedghostimaging}, or use compressive sensing to speed-up measurement~\cite{compressiveghostsensing}.
Furthermore, as long as the patterns used are known, they do not need to be deterministically generated or even orthogonal, and even a set of speckle patterns allow to reconstruct an image~\cite{speckleghostimaging}.

A property that is shared by all variants of ghost imaging is that one needs to know exactly what the set of illumination patterns is. What is effectively measured with the single pixel detector is proportional to the overlap between the object $O$ and the illumination pattern $P_i$, i.e. the coefficient $b_i = \int P_i (\mathbf{r}) O(\mathbf{r}) d\mathbf{r}$. If the set of illumination patterns forms a complete basis, one can reconstruct an image of the object as $I(\mathbf{r}')= \sum_i b_i P_i(\mathbf{r}')$, but if the patterns $P_i$ are unknown, this approach breaks down.

In this article we show that, even if the illumination patterns are completely unknown, one can still use a different set of patterns in the reconstruction formula, as long as this second set is correlated with the first one. In particular, we exploit the recently discovered spatial correlation between the transmitted and reflected speckle patterns generated at both sides of a scattering medium~\cite{CRtheory1,CRTcorr}, which allows us to reconstruct the shape of an object hidden behind a turbid medium, potentially fully opaque, using only the reflected speckle pattern, instead of the transmitted one. Furthermore, we generalize this technique to a completely non-invasive geometry, where both the  camera measuring the speckle pattern and the single-pixel detector are on the same side of the scattering layer, allowing to image a fluorescent object  placed on the other side.
\begin{figure}[tb]
	\centering
	\includegraphics[width=1\linewidth]{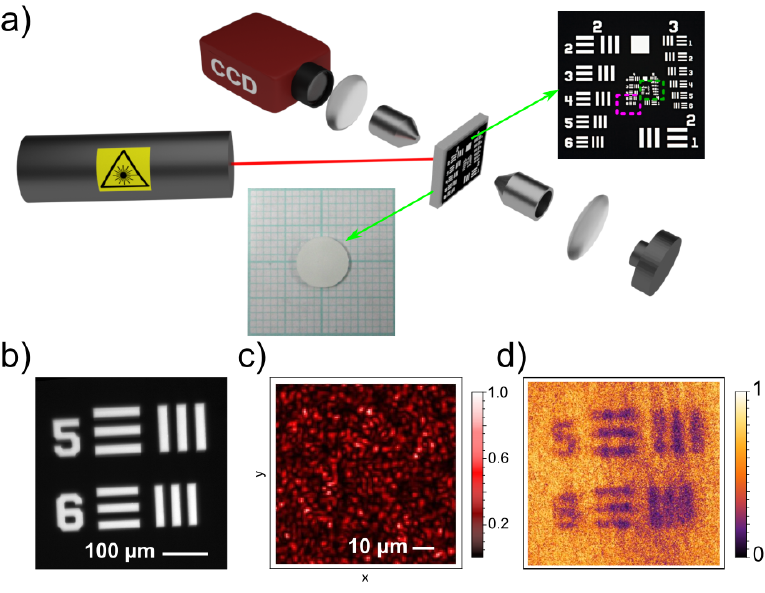}
	\caption{a) Experimental apparatus. A cw laser illuminates an opaque scattering material and an object hidden behind (insets). An imaging system records the reflected speckle pattern from the surface of the scattering sample and a bucket detector collects the intensity transmitted by the object. b) Elements 5 and 6 of Group 4 of the resolution target used as object to image in this experiment, highlighted by the pink square in the inset of panel a. c) Typical speckle pattern collected in reflection with the imaging system presented. d) Retrieved image using BGI with $2.27 \times10^6$ disorder realizations. }
	\label{fig:pic1}
\end{figure}

When using speckle to perform ghost imaging, one usually sends a laser beam through a time-varying scattering medium, often a rotating diffuser, and the resulting transmitted intensity speckle pattern, $T_i$, is measured and used as the illumination pattern $P_i$. The transmitted light passing through the object is then integrated and measured with a single pixel detector, yielding the coefficient $b_i= \int T_i (\mathbf{r}) O(\mathbf{r}) d\mathbf{r}$. Full knowledge of both $b_i$ and $T_i$ allows one, for a large enough number $N$ of patterns, to obtain a faithful representation of the object $O$.  In order to measure directly the transmitted speckle patterns, one needs to have an imaging system placed behind the scattering layer. In most practical situations, this is actually not possible, e.g. because access is restricted, as in biomedical imaging. In these cases one can  rely on the reflected speckle patterns $R_i$ only, which share mutual information with the transmitted ones in the form of a spatial correlation~\cite{CRtheory1,CRTcorr,CRtheory2}. The simplest approach we can take to make use of this mutual information is to replace each $T_i (\mathbf{r}')$ with $R_i(\mathbf{r}')$, which results in the reconstructed image
\begin{equation}
 \tilde{I}(\mathbf{r}')=\sum_{i=1}^N b_i R_i(\mathbf{r}').
 \label{eq:bgiprocedure}
 \end{equation}
Identifying the sum $\sum_{i=1}^N(\dots)_i$ with the ensemble average $\langle \dots \rangle$ and substituting in the definition of $b_i$, we can express the reconstructed image as 
\begin{equation}
\begin{aligned}
 \tilde{I}(\mathbf{r}') =&  \langle \int O(\mathbf{r}) T (\mathbf{r}) R (\mathbf{r}') d\mathbf{r} \rangle  \\
 =& \int O(\mathbf{r}) \langle  T (\mathbf{r}) R (\mathbf{r}') \rangle d\mathbf{r} \\
 =& \langle T \rangle \langle R \rangle \left[ O \ast C^{RT} + \int O(\mathbf{r}) d\mathbf{r} \right] \\
 \propto& \, O \ast C^{RT} + \mathcal{A}
 \label{eq:resolution}
\end{aligned}
\end{equation}
where $C^{RT}(\Delta \mathbf{r})=\langle  \delta R (\mathbf{r}) \delta T (\mathbf{r}+\Delta \mathbf{r})\rangle$ is the normalized correlation function of the reflected and transmitted intensity patterns ($\delta f=f/\langle f \rangle -1$ denotes the normalized statistical fluctuation of the random variable $f$) and the constant $ \mathcal{A}= \int O(\mathbf{r})d\mathbf{r}$ represents a flat background proportional to the total signal from the object. Hence, using the reflected speckle patterns instead of the transmitted ones, we obtain the very same image, but with a lower resolution, given by the range of the correlation function $C^{RT}$, which acts as a point spread function. We name this Blind Ghost Imaging (BGI), as it allows to perform ghost imaging without ever knowing the patterns used to illuminate the object.\\ 

To verify our prediction we designed an experiment where we image an object hidden behind an opaque scattering medium. The experimental apparatus is shown in Fig.~\ref{fig:pic1}a. A \SI{2}{\milli\watt} He-Ne laser is incident on a scattering medium (Fig.~\ref{fig:pic1}a, inset) at an angle of approximately 45$^\circ$ with respect to the sample surface. In this way, contributions of the specularly reflected and ballistically transmitted light, which spoil the correlation  $C^{RT}$, are not collected~\cite{CRTcorr}. The scattering layer is made of a suspension of TiO$_2$ particles in glycerol, with a scattering mean free path $\ell=\SI{16}{} \pm \SI{2}{\micro\meter}$ and a \SI{40}{\micro\meter} thickness, resulting in an Optical Density ($OD$) $\simeq 2.5$. The object to image, a Thorlabs USAF 1951 calibration test target (Fig.~\ref{fig:pic1}b), is in contact with the scattering layer. The reflected speckle pattern (Fig.~\ref{fig:pic1}c) is imaged on the scattering medium surface and recorded using a CCD camera. As the scattering layer is liquid, the speckle patterns change with time which allows us to record a large number of different speckle patterns without moving or changing the sample. The transmitted light passing through the object is then integrated by a bucket detector. For simplicity of alignment, this is done by using  an identical CCD camera and integrating over all pixels. This allows us to measure the correlation $C^{RT}(\Delta \mathbf{r})$, discussed later on, using the same apparatus. 

In Fig.~\ref{fig:pic1}d we show the reconstructed image of the object represented in Fig.~\ref{fig:pic1}b, when using the reflected speckle patterns and integrating the transmitted intensity, according to Eq.~(\ref{eq:bgiprocedure}). Here, we used $N=2.27 \times10^6$ realizations of the disorder. Apart from the residual noise, the object is clearly visible and all features are resolved. We notice that a gaussian smoothing of the picture would remove most of the noise, producing a more pleasing image.
This experiment demonstrates that it is possible to perform ghost imaging using a set of patterns different from the illuminating one but correlated with it. In particular it is possible to use the reflected, instead of the transmitted speckle, to reconstruct the shape of an object placed behind an opaque scattering layer. Compared to other ghost imaging schemes using reflected signal~\cite{GIreflective,GIbackscattering}, this method works in the deep multiple scattering regime without making use of any ballistic light.
\begin{figure}[tb]
	\centering
	\includegraphics[width=1\linewidth]{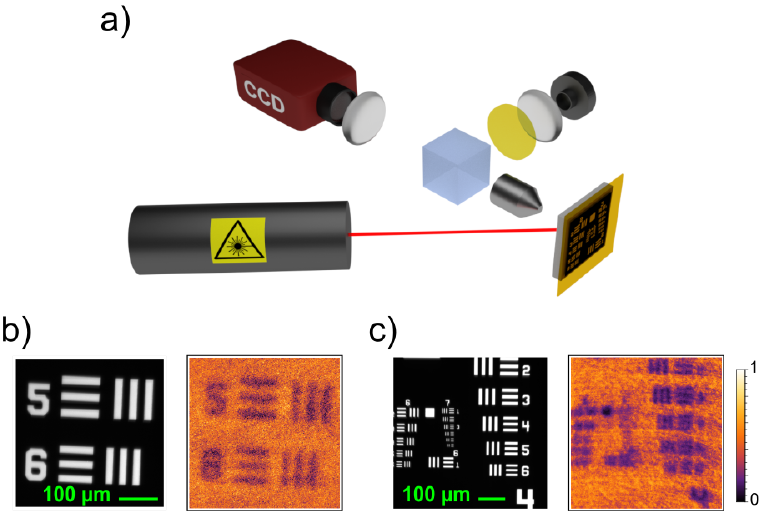}
	\caption{ 
	 a) Experimental apparatus used for non-invasive BGI. A \SI{450}{nm} laser is incident on the scattering sample at $\approx 45^\circ$. The resolution target is placed on the back surface of the scattering material, and right behind it we have a fluorescent layer (Cerium-doped YAG), acting as a fluorescent object. The bucket detector is in this case also in reflection from the sample, filtering the fluorescent light with a \SI{500}{nm} long pass filter. b)  Elements 5 and 6 of Group 4 of the resolution target used as the object, and the image retrieved using BGI with $4\times10^6$ disorder realizations. c)  Object representing Groups 5, 6 and 7 from the resolution target, and the image retrieved using BGI with $1.5 \times10^6$ disorder realizations. }
	\label{fig:pic2}
\end{figure}

As the bucket detector does not have any spatial resolution, there is no fundamental reason why it should be placed behind the object as in traditional ghost imaging. This suggest that blind ghost imaging can be adapted to a completely non-invasive configuration. We modified the apparatus so that all optical components are on the opposite side of the scattering layer with respect to the object, as shown in Fig.~\ref{fig:pic2}a. The fluorescent sample consists of the USAF negative target with a fluorescent layer of Cerium-doped YAG just behind it. The illumination geometry is the same as in the first experiment, but in this case we used a \SI{100}{\milli\watt} blue laser (\SI{450}{\nano\meter}) producing a white fluorescent emission from the Cerium-doped YAG layer. Both the reflected speckle and the fluorescence are collected by a 10x microscope objective, and a plano-convex \SI{150}{mm} lens, in an epi configuration. The speckle pattern is recorded by a CCD camera, and the fluorescence from the object is collected by the bucket detector after passing through a long-pass \SI{500}{\nano\meter} filter. Again in this case the bucket detector is a CCD with the intensity integrated over all pixels. In Fig.~\ref{fig:pic2}b we show the retrieved image for this case, obtained with $N=4 \times10^6$ disorder realizations. The image is very well reconstructed, with an outcome very similar to the one shown in Fig.~\ref{fig:pic1}d obtained with the bucket placed on the transmission side.\\ 

\begin{figure}[bt]
	\centering
	\includegraphics[width=1\linewidth]{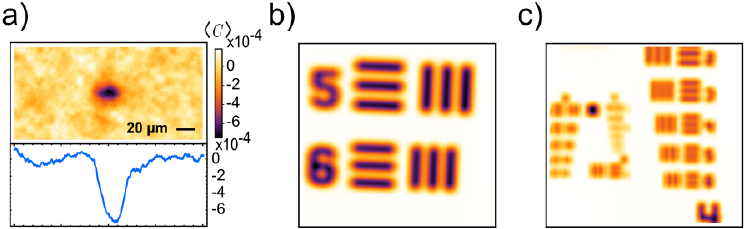}
	\caption{ a) 2D map and a 1D cross section along $\Delta y=0$ of the averaged correlation between the transmitted and reflected speckle patterns. b) and c) Expected images obtained by numerically convolving the objects shown in Fig.~\ref{fig:pic2}b,c with the correlation function shown in a). 
}
	\label{fig:pic3}
\end{figure}

In order to evaluate the performances of the blind ghost imaging setup, we first took an image of the elements 5 and 6 of group 4 of the USAF target, as shown in Fig.~\ref{fig:pic2}b, and found a lateral resolution $\Delta r \simeq$ \SI{20}{\micro\meter}. We then repeated the measurement with an object with smaller features (groups 5, 6 and 7 of the resolution target) shown in Fig.~\ref{fig:pic2}c, in order to better quantify the resolution of this method. According to our prediction~(Eq.~\ref{eq:resolution}), this resolution should be given by the width of the correlation $C^{RT}(\Delta \mathbf{r})$ which acts as a point spread function. To confirm that this is indeed the case, we made a separate measurement of the average intensity correlation between transmitted and reflected speckle patterns~\cite{CRTcorr}, and compared the blind ghost imaging results of Fig.~\ref{fig:pic2}b,c with the numerical convolution of the object with $C^{RT}$. Results are presented in Fig.~\ref{fig:pic3}. The retrieved images  (Fig.~\ref{fig:pic3}b,c) resemble very well the expected ones (Fig.~\ref{fig:pic2}b,c), resolving the same elements and thus demonstrating that the resolution of the resulting image depends on the width and shape of the correlation function $C^{RT}$ (Fig.~\ref{fig:pic3}a), as dictated by Eq.~(\ref{eq:resolution}). In particular, the width of the correlation function limits the features of the object that can be resolved, even in the ideal and noise-free case, where it is possible to resolve mainly the first few elements of group 5. 

The shape and the sign of the correlation $C^{RT}$ depend both on the sample thickness $L$ and the transport mean free path $\ell$ in a non-trivial way~\cite{CRTcorr}. However, in the multiple scattering regime ($L\gtrsim\ell$), it takes a simple form, mostly isotropic and negative, with a width $\sim L$, as shown in Fig.~\ref{fig:pic3}a. The negative sign of the correlation is the reason why the images appear as a negative signal on top of a bright background. In addition, the width scaling can be understood from the microscopic scattering process responsible for the correlation~\cite{Feng1988,stephen1987intensity,vanrossum99}. Interferences between scattered waves create a bulk speckle pattern inside the disordered medium, which acts as an ensemble of local fluctuating sources for diffusive transport~\cite{zyuzin1987langevin,pnini1989fluctuations,van1990observation,genack90,de1992transmission}. Two diffusive paths generated by the same source and emerging on opposite sides of the sample are thus correlated~\cite{rogozkin95,froufe2007fluctuations}. Since diffusive paths explore a domain of transverse size bounded by $L$, the range of $C^{RT}$ necessarily scales linearly with $L$.  This means that the resolution of the blind ghost imaging scheme is given by the depth of the target object. This spatial resolution is comparable to that obtained in diffuse optical imaging, which uses a CCD camera in transmission instead of a simple bucket detector~\cite{opticaldiffuseimaging}. 

Another specific feature of the blind ghost imaging scheme is its signal to noise ratio (SNR), which depends on the amplitude and the range of the correlation $C^{RT}$, as well as the size of the illuminated object. As discussed above, $C^{RT}$ has a width of order $L$ and a small amplitude $\alpha$, so that the useful signal (i.e. first term of Eq.~(\ref{eq:resolution})) is always smaller than the constant background $\mathcal{A}$. In addition, because of the Rayleigh-like statistics of the speckle patterns used to reconstruct the image, fluctuations are large and proportional to the full signal. This results in a SNR $\sim \sqrt{N}\alpha L^2/\mathcal{A}$ (see SI for details). Typically in our experiment $\alpha\sim10^{-3}$, which imposes a number of measurements $N\gtrsim 10^6$ to get SNR $\gtrsim 1$. In the deep diffusive regime, $L\gg \ell$, which is not reached in our experiment, it is known that $\alpha\sim\lambda^2/L^2$~\cite{CRTcorr}, so that the correlation $C^{RT}$ becomes independent of the disorder strength parametrized by the mean free path $\ell$, and the SNR  independent of both $L$ and $\ell$ (SNR $\sim \sqrt{N} \lambda^2/\mathcal{A}$). This analysis shows that blind ghost imaging can, in principle, be used to take the image of an object hidden behind a fully opaque medium in the deep diffusive regime.  

\begin{figure}[tb]
	\centering
	\includegraphics[width=1\linewidth]{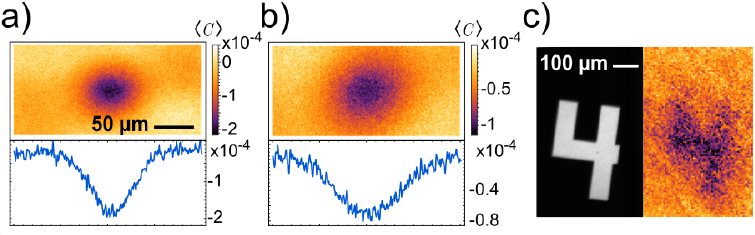}
	\caption{ a) and b) Correlation functions between the reflected and transmitted speckle patterns measured \SI{80}{\micro\meter} and \SI{160}{\micro\meter} respectively away from the transmission and reflection surfaces. c) Object separated by a cover slip of \SI{150}{\micro\meter} from the scattering medium and retrieved image using BGI with $5.65 \times10^5$ disorder realizations.}
	\label{fig:pic4}
\end{figure}

In the experiments described above, the object to be imaged was placed right on the back of the scattering layer and the reflected speckle pattern was recorded at its front surface. In this configuration, $C^{RT}$ is expected to be maximally peaked~\cite{CRTcorr}. Since the latter originates from bulk speckle patterns, and thus from interferences, we could wonder how $C^{RT}$ is modified when the object is further away from the surface.
As free space propagation preserve mutual information, the integral of $C^{RT}(\Delta \mathbf{r})$ must be constant even when it is measured between two planes away from the scattering layer. At the same time, we expect that the mutual information will spread over larger and larger areas, until it becomes a constant function in the far field. To be more quantitative, we extended the theoretical analysis of Ref.~\cite{CRTcorr} and computed analytically $C^{RT}$ on two planes at arbitrary distances, $D$ and $D'$, away from the sample. We found that, in the regime $L\gg \ell$, one obtains the simple form $C^{RT}(\Delta \mathbf{r}, D, D')=C^{RT}(\Delta \mathbf{r}, 0,0)\ast h(\Delta \mathbf{r}, D) \ast h(\Delta \mathbf{r}, D')$, where $h(\Delta \mathbf{r}, D)$ is a normalized function of width $\sim D$ (see SI for details). This means that objects located further away from the scattering layer can be imaged with almost unaffected resolution and contrast as long as $D,D' \ll L$. It also implies that the image quality does not depend on the exact position of the disordered sample, but rather on the distance between the object and the plane where the reflected speckle is imaged. To test these predictions, we measured the correlation $C^{RT}$ from the same sample used in the previous experiments, on two planes placed at various distances from the sample. Representative results are shown in Fig.~\ref{fig:pic4}a,b for planes at \SI{80}{\micro\meter} and \SI{160}{\micro\meter} respectively away from the sample (see SI for a systematic study). As  can be seen, the correlation becomes indeed wider, but does so gradually. Hence, it is possible to use blind ghost imaging to image objects away from the scattering layer at the price of a reduced resolution, but without introducing complicated aberrations. This is illustrated in Fig.~\ref{fig:pic4}c, where we show an object and its blind ghost imaging retrieved image, when the reflected speckle pattern was measured on the surface of the sample and the scattering medium is \SI{150}{\micro\meter} away from the object. The number of measurements needed to retrieve that image was $N=1.5 \times10^5$. This experiment successively mimics a situation where one does not necessarily know how far away the object is from the scattering layer.\\

In conclusion, we have demonstrated ghost imaging through an opaque scattering medium without measuring the transmitted speckle pattern that illuminates the target. This blind ghost imaging scheme uses instead a measurement of the reflected speckle, that is merely spatially correlated with the transmitted one. The achievable resolution is given by the width of the correlation function, while the number of realizations of the disorder needed to obtain a noiseless image depends both on the amplitude of the correlation function and the total signal received by the bucket detector. Fundamentally, our results illustrate an important feature of ghost imaging, namely, that one does not need to measure the illuminating signal, but only a signal weakly correlated to it. Practically, this broadens the potential range of applications of ghost imaging, in particular for non-invasive imaging in biological tissues. 
Several possible strategies can be used to improve the processing speed, limited by the large amount of measurements required to reach a viable signal to noise ratio: fast-moving scattering media in conjunction with fast cameras will naturally reduce measurement time, but for slow-moving media one can generate different (unknown) illumination patterns by modulating the incident wavefront with a spatial light modulator. Alternatively compressive sensing techniques could reduce the number of necessary measurements, as long as some assumption (e.g. sparsity) can be made about the object to be imaged~\cite{compressiveghostsensing}.

\section*{Methods}
The scattering medium is made of a suspension of TiO$_2$ particles in glycerol with a concentration of $\SI{300}{\milli g} $ of TiO$_2$ for $\SI{10}{\milli\liter}$ of glycerol, which lead to a scattering mean free path $\ell=\SI{16}{} \pm \SI{2}{\micro\meter}$. The suspension is held between one glass slide and the resolution target that works as the object to image, and its thickness is controlled using calibrated feeler gauges. Throughout  the experiments described here we used a fixed $L=\SI{40}{\micro\meter}$ thickness.

The reflected speckle pattern (see Fig.~\ref{fig:pic1}c for a typical measurement) is recorded at the surface of the scattering medium using a conventional imaging system, composed of a 10x microscope objective, a plano-convex \SI{150}{\milli\meter} lens and a CCD camera (Allied Vision Manta G-146). As glycerol is very viscous, we used a piezoelectric buzzer attached to the glass slide holding the sample to speed up the movement of the particles and shorten the decorrelation time of the generated speckle patterns, which allowed us to record different speckle patterns at the maximal acquisition speed of the cameras, around 17 frames per second, and thus to perform an ensemble average.

\bibliography{bibliography}

\end{document}